\documentclass[letterpaper,aps,prd,superscriptaddress,showpacs,nofootinbib,floatfix,twocolumn]{revtex4}

\usepackage{graphicx}	

\usepackage{xspace}	

\def\gev{\mbox{~GeV}}
\def\gevc{\mbox{~GeV/$c$}}

\def\mevc{\mbox{~MeV/$c$}}

\newcommand{\pp} {\ensuremath{p+p}}

\newcommand{\ee} {\ensuremath{e^+ e^-}}
\newcommand{\pt} {\ensuremath{p_T}}

\def\la{\left< }
\def\ra{\right> }
\def\mean#1{\ensuremath{\la#1\ra}}
 
\newcommand{\ptt}{\ensuremath{p_{T\rm t}}}
 
\newcommand{\pta}{\ensuremath{p_{T\rm a}}}

\newcommand{\pout} {\ensuremath{p_{\rm out}}}

\newcommand{\all} {\ensuremath{A_{LL}}}
\newcommand{\kt} {\ensuremath{k_T}}
\newcommand{\ktrms} {\ensuremath{\sqrt{\langle \kt ^2 \rangle} }}
\newcommand{\jtrms} {\ensuremath{\sqrt{\langle \jt ^2 \rangle} }}

\newcommand{\jt} {\ensuremath{j_T}}

\newcommand{\zt} {\ensuremath{z_{\rm t}}}
\newcommand{\mzt} {\mean{\zt}}

\newcommand{\xhh} {\ensuremath{\hat{x}_{\rm h}}}

\newcommand{\s} {\ensuremath{\sqrt{s}}}
\newcommand{\piz} {\ensuremath{\pi^0}}
\newcommand{\pizh}{\ensuremath{\pi^0-h^\pm}}

\begin{document}


\title{Double-Helicity Dependence of Jet Properties from Dihadrons in 
Longitudinally Polarized $p+p$~Collisions at 
$\sqrt{s}$~=~200~GeV}

\newcommand{\abilene}{Abilene Christian University, Abilene, TX 79699, U.S.}
\newcommand{\acadsin}{Institute of Physics, Academia Sinica, Taipei 11529, Taiwan}
\newcommand{\banaras}{Department of Physics, Banaras Hindu University, Varanasi 221005, India}
\newcommand{\barc}{Bhabha Atomic Research Centre, Bombay 400 085, India}
\newcommand{\bnlcoll}{Collider-Accelerator Department, Brookhaven National Laboratory, Upton, NY 11973-5000, U.S.}
\newcommand{\bnlphys}{Physics Department, Brookhaven National Laboratory, Upton, NY 11973-5000, U.S.}
\newcommand{\caucr}{University of California - Riverside, Riverside, CA 92521, U.S.}
\newcommand{\charlesczech}{Charles University, Ovocn\'{y} trh 5, Praha 1, 116 36, Prague, Czech Republic}
\newcommand{\ciae}{China Institute of Atomic Energy (CIAE), Beijing, People's Republic of China}
\newcommand{\cns}{Center for Nuclear Study, Graduate School of Science, University of Tokyo, 7-3-1 Hongo, Bunkyo, Tokyo 113-0033, Japan}
\newcommand{\colorado}{University of Colorado, Boulder, CO 80309, U.S.}
\newcommand{\columbia}{Columbia University, New York, NY 10027 and Nevis Laboratories, Irvington, NY 10533, U.S.}
\newcommand{\czechtech}{Czech Technical University, Zikova 4, 166 36 Prague 6, Czech Republic}
\newcommand{\dapnia}{Dapnia, CEA Saclay, F-91191, Gif-sur-Yvette, France}
\newcommand{\debrecen}{Debrecen University, H-4010 Debrecen, Egyetem t{\'e}r 1, Hungary}
\newcommand{\elte}{ELTE, E{\"o}tv{\"o}s Lor{\'a}nd University, H - 1117 Budapest, P{\'a}zm{\'a}ny P. s. 1/A, Hungary}
\newcommand{\fit}{Florida Institute of Technology, Melbourne, FL 32901, U.S.}
\newcommand{\fsu}{Florida State University, Tallahassee, FL 32306, U.S.}
\newcommand{\gsu}{Georgia State University, Atlanta, GA 30303, U.S.}
\newcommand{\hiroshima}{Hiroshima University, Kagamiyama, Higashi-Hiroshima 739-8526, Japan}
\newcommand{\ihepprot}{IHEP Protvino, State Research Center of Russian Federation, Institute for High Energy Physics, Protvino, 142281, Russia}
\newcommand{\illuiuc}{University of Illinois at Urbana-Champaign, Urbana, IL 61801, U.S.}
\newcommand{\instpasczech}{Institute of Physics, Academy of Sciences of the Czech Republic, Na Slovance 2, 182 21 Prague 8, Czech Republic}
\newcommand{\isu}{Iowa State University, Ames, IA 50011, U.S.}
\newcommand{\jinrdubna}{Joint Institute for Nuclear Research, 141980 Dubna, Moscow Region, Russia}
\newcommand{\kek}{KEK, High Energy Accelerator Research Organization, Tsukuba, Ibaraki 305-0801, Japan}
\newcommand{\kfki}{KFKI Research Institute for Particle and Nuclear Physics of the Hungarian Academy of Sciences (MTA KFKI RMKI), H-1525 Budapest 114, POBox 49, Budapest, Hungary}
\newcommand{\korea}{Korea University, Seoul, 136-701, Korea}
\newcommand{\kurchatov}{Russian Research Center ``Kurchatov Institute", Moscow, Russia}
\newcommand{\kyoto}{Kyoto University, Kyoto 606-8502, Japan}
\newcommand{\labllr}{Laboratoire Leprince-Ringuet, Ecole Polytechnique, CNRS-IN2P3, Route de Saclay, F-91128, Palaiseau, France}
\newcommand{\lawllnl}{Lawrence Livermore National Laboratory, Livermore, CA 94550, U.S.}
\newcommand{\losalamos}{Los Alamos National Laboratory, Los Alamos, NM 87545, U.S.}
\newcommand{\lpc}{LPC, Universit{\'e} Blaise Pascal, CNRS-IN2P3, Clermont-Fd, 63177 Aubiere Cedex, France}
\newcommand{\lund}{Department of Physics, Lund University, Box 118, SE-221 00 Lund, Sweden}
\newcommand{\mass}{Department of Physics, University of Massachusetts, Amherst, MA 01003-9337, U.S. }
\newcommand{\muenster}{Institut f\"ur Kernphysik, University of Muenster, D-48149 Muenster, Germany}
\newcommand{\muhlenberg}{Muhlenberg College, Allentown, PA 18104-5586, U.S.}
\newcommand{\myongji}{Myongji University, Yongin, Kyonggido 449-728, Korea}
\newcommand{\nagasaki}{Nagasaki Institute of Applied Science, Nagasaki-shi, Nagasaki 851-0193, Japan}
\newcommand{\newmex}{University of New Mexico, Albuquerque, NM 87131, U.S. }
\newcommand{\nmsu}{New Mexico State University, Las Cruces, NM 88003, U.S.}
\newcommand{\ornl}{Oak Ridge National Laboratory, Oak Ridge, TN 37831, U.S.}
\newcommand{\orsay}{IPN-Orsay, Universite Paris Sud, CNRS-IN2P3, BP1, F-91406, Orsay, France}
\newcommand{\peking}{Peking University, Beijing, People's Republic of China}
\newcommand{\pnpi}{PNPI, Petersburg Nuclear Physics Institute, Gatchina, Leningrad region, 188300, Russia}
\newcommand{\riken}{RIKEN Nishina Center for Accelerator-Based Science, Wako, Saitama 351-0198, JAPAN}
\newcommand{\rikjrbrc}{RIKEN BNL Research Center, Brookhaven National Laboratory, Upton, NY 11973-5000, U.S.}
\newcommand{\rikkyo}{Physics Department, Rikkyo University, 3-34-1 Nishi-Ikebukuro, Toshima, Tokyo 171-8501, Japan}
\newcommand{\saispbstu}{Saint Petersburg State Polytechnic University, St. Petersburg, Russia}
\newcommand{\saopaulo}{Universidade de S{\~a}o Paulo, Instituto de F\'{\i}sica, Caixa Postal 66318, S{\~a}o Paulo CEP05315-970, Brazil}
\newcommand{\seoulnat}{System Electronics Laboratory, Seoul National University, Seoul, Korea}
\newcommand{\stonybrkc}{Chemistry Department, Stony Brook University, Stony Brook, SUNY, NY 11794-3400, U.S.}
\newcommand{\stonycrkp}{Department of Physics and Astronomy, Stony Brook University, SUNY, Stony Brook, NY 11794, U.S.}
\newcommand{\subatech}{SUBATECH (Ecole des Mines de Nantes, CNRS-IN2P3, Universit{\'e} de Nantes) BP 20722 - 44307, Nantes, France}
\newcommand{\tenn}{University of Tennessee, Knoxville, TN 37996, U.S.}
\newcommand{\titech}{Department of Physics, Tokyo Institute of Technology, Oh-okayama, Meguro, Tokyo 152-8551, Japan}
\newcommand{\tsukuba}{Institute of Physics, University of Tsukuba, Tsukuba, Ibaraki 305, Japan}
\newcommand{\vandy}{Vanderbilt University, Nashville, TN 37235, U.S.}
\newcommand{\waseda}{Waseda University, Advanced Research Institute for Science and Engineering, 17 Kikui-cho, Shinjuku-ku, Tokyo 162-0044, Japan}
\newcommand{\weizmann}{Weizmann Institute, Rehovot 76100, Israel}
\newcommand{\yonsei}{Yonsei University, IPAP, Seoul 120-749, Korea}
\affiliation{\abilene}
\affiliation{\acadsin}
\affiliation{\banaras}
\affiliation{\barc}
\affiliation{\bnlcoll}
\affiliation{\bnlphys}
\affiliation{\caucr}
\affiliation{\charlesczech}
\affiliation{\ciae}
\affiliation{\cns}
\affiliation{\colorado}
\affiliation{\columbia}
\affiliation{\czechtech}
\affiliation{\dapnia}
\affiliation{\debrecen}
\affiliation{\elte}
\affiliation{\fit}
\affiliation{\fsu}
\affiliation{\gsu}
\affiliation{\hiroshima}
\affiliation{\ihepprot}
\affiliation{\illuiuc}
\affiliation{\instpasczech}
\affiliation{\isu}
\affiliation{\jinrdubna}
\affiliation{\kek}
\affiliation{\kfki}
\affiliation{\korea}
\affiliation{\kurchatov}
\affiliation{\kyoto}
\affiliation{\labllr}
\affiliation{\lawllnl}
\affiliation{\losalamos}
\affiliation{\lpc}
\affiliation{\lund}
\affiliation{\mass}
\affiliation{\muenster}
\affiliation{\muhlenberg}
\affiliation{\myongji}
\affiliation{\nagasaki}
\affiliation{\newmex}
\affiliation{\nmsu}
\affiliation{\ornl}
\affiliation{\orsay}
\affiliation{\peking}
\affiliation{\pnpi}
\affiliation{\riken}
\affiliation{\rikjrbrc}
\affiliation{\rikkyo}
\affiliation{\saispbstu}
\affiliation{\saopaulo}
\affiliation{\seoulnat}
\affiliation{\stonybrkc}
\affiliation{\stonycrkp}
\affiliation{\subatech}
\affiliation{\tenn}
\affiliation{\titech}
\affiliation{\tsukuba}
\affiliation{\vandy}
\affiliation{\waseda}
\affiliation{\weizmann}
\affiliation{\yonsei}
\author{A.~Adare} \affiliation{\colorado}
\author{S.~Afanasiev} \affiliation{\jinrdubna}
\author{C.~Aidala} \affiliation{\columbia}  \affiliation{\losalamos} \affiliation{\mass} 
\author{N.N.~Ajitanand} \affiliation{\stonybrkc}
\author{Y.~Akiba} \affiliation{\riken} \affiliation{\rikjrbrc}
\author{H.~Al-Bataineh} \affiliation{\nmsu}
\author{J.~Alexander} \affiliation{\stonybrkc}
\author{K.~Aoki} \affiliation{\kyoto} \affiliation{\riken}
\author{L.~Aphecetche} \affiliation{\subatech}
\author{R.~Armendariz} \affiliation{\nmsu}
% ...
\collaboration{PHENIX Collaboration} \noaffiliation

\date{\today}

\begin{abstract}

It has been postulated that partonic orbital angular momentum can 
lead to a significant double-helicity dependence in the net 
transverse momentum of Drell-Yan dileptons produced in 
longitudinally polarized $p+p$ collisions.  Analogous effects are 
also expected for dijet production. If confirmed by experiment, 
this hypothesis, which is based on semi-classical arguments, 
could lead to a new approach for studying the contributions of 
orbital angular momentum to the proton spin. We report the first 
measurement of the double-helicity dependence of the dijet 
transverse momentum in longitudinally polarized $p+p$ collisions 
at $\sqrt{s}$~=~200 GeV from data taken by the PHENIX 
experiment in 2005 and 2006.  The analysis deduces the transverse 
momentum of the dijet from the widths of the near- and far-side 
peaks in the azimuthal correlation of the dihadrons.  When 
averaged over the transverse momentum of the triggered particle, 
the difference of the root-mean-square of the dijet transverse 
momentum between like- and unlike-helicity collisions is found to 
be $-37 \pm 88^{\rm stat} \pm 14^{\rm syst}$ MeV/$c$.

\end{abstract}

	

\maketitle

\section{Introduction \label{sec:Intro}}

Since the startling 1989 result of the European Muon 
Collaboration, which revealed that much less of the proton spin 
is carried by the quark and antiquark spins than previously 
expected~\cite{ref:EMC}, there has been great interest in the 
angular momentum structure of the nucleon.  Subsequent 
deep-inelastic scattering experiments have confirmed that only 
$\sim$20-30\% of the proton spin is due to quark and antiquark 
polarization~\cite{ref:HERMES6,ref:COMPASS}.

The remainder of the spin of the proton must be due to gluon spin 
and/or partonic orbital angular momentum (OAM). It is known that 
the proton anomalous magnetic moment requires orbital angular 
momentum of the quarks, although the combined orbital angular 
momentum of all flavors may be close to zero (see e.g. 
\cite{ref:BURKARDTSCHNELL,Avakian:2007xa}). Recent measurements 
of $\Delta G$, the gluon spin contribution to the proton, are 
still statistically limited but have excluded large values of 
gluon polarization 
\cite{ref:PHENIX_pi0_all3,ref:STARjet,ref:PHENIX_pi0_all4}, and 
the most recent global study indicates nearly vanishing gluon 
polarization in the presently accessible $x$ range, together with 
a small quark polarization~\cite{deFlorian:2009vb}. 
Forthcoming 
data from the Relativistic Heavy Ion Collider (RHIC) should place 
tighter constraints on $\Delta G$ and shed new light on the spin 
puzzle. Meanwhile, progress in the quark and gluon helicity 
distributions has served to help fuel the increasing interest in 
orbital angular momentum that began in the 1990s.

It is important to note that while the total spin of the proton 
as $\frac{1}{2}\hbar$ is well defined, there is no unique way to 
describe the decomposition of the angular momentum among the 
interacting partons within a nucleon (see e.g. 
\cite{ref:Burkardte}). Thus discussions of partonic orbital 
angular momentum in the proton typically involve a number of 
subtleties, despite the relatively intuitive nature of the 
concept. Two decompositions of nucleon angular momentum that have 
become standard are that of Jaffe and Manohar 
\cite{ref:JaffeSumRule} and that of Ji~\cite{ref:JI}. While at 
present no quantitative method is known to probe experimentally 
the partonic OAM of the Jaffe-Manohar decomposition, in Ji's 
paper he proposes the experimental technique of deeply-virtual 
Compton scattering to access quark OAM via Generalized Parton 
Distributions (GPDs). Several groups have already pursued this 
experimentally challenging path 
\cite{MunozCamacho:2006hx,Airapetian:2006zr,Mazouz:2007vj,Girod:2007jq,Airapetian:2008jga}. 
Initial measurements of hard exclusive leptoproduction of 
vector mesons, another means of accessing GPDs, have also 
been performed~\cite{Morrow:2008ek,Airapetian:2009ut}.  
Within the Ji decomposition, results for the OAM of up and 
down quarks have become available from lattice quantum 
chromodynamics (QCD) calculations~\cite{ref:lattice}.  These 
lattice QCD results suggest that the orbital angular momentum 
for $u$ and $d$ quarks separately is quite substantial, but 
that these contributions largely cancel in the proton.

Another approach to studying the transverse motion of quarks 
and gluons within the nucleon is through 
transverse-momentum-dependent parton distribution functions 
(TMDs).  The first attempt to use a TMD to describe the large 
transverse single-spin asymmetries (SSAs) observed in 
polarized hadronic collisions was made by Sivers in a 1990 
paper~\cite{ref:Sivers}, and the various TMDs contributing to 
the leading-order polarized semi-inclusive deep-inelastic 
scattering (DIS) cross section were laid out by Mulders and 
Tangerman in 1996~\cite{Mulders:1995dh}.  Progress was made 
in both experiment and theory throughout the decade, but it 
was only after some key theoretical developments in 2002-03 
\cite{ref:Brodsky,Collins:2002kn,Belitsky:2002sm} that an 
ongoing period of intense theoretical and experimental 
activity regarding TMDs began. It should be noted that thus 
far, no model-independent quantitative relationship between 
TMDs and parton orbital angular momentum has been derived 
\cite{ref:Meissner,Burkardt:2009sh}, and it is not clear at 
present if the OAM to which TMDs could provide sensitivity 
would fit within either the Jaffe-Manohar or Ji decomposition 
of nucleon angular momentum.

\begin{figure}[tb]
\includegraphics[width=0.48\linewidth]{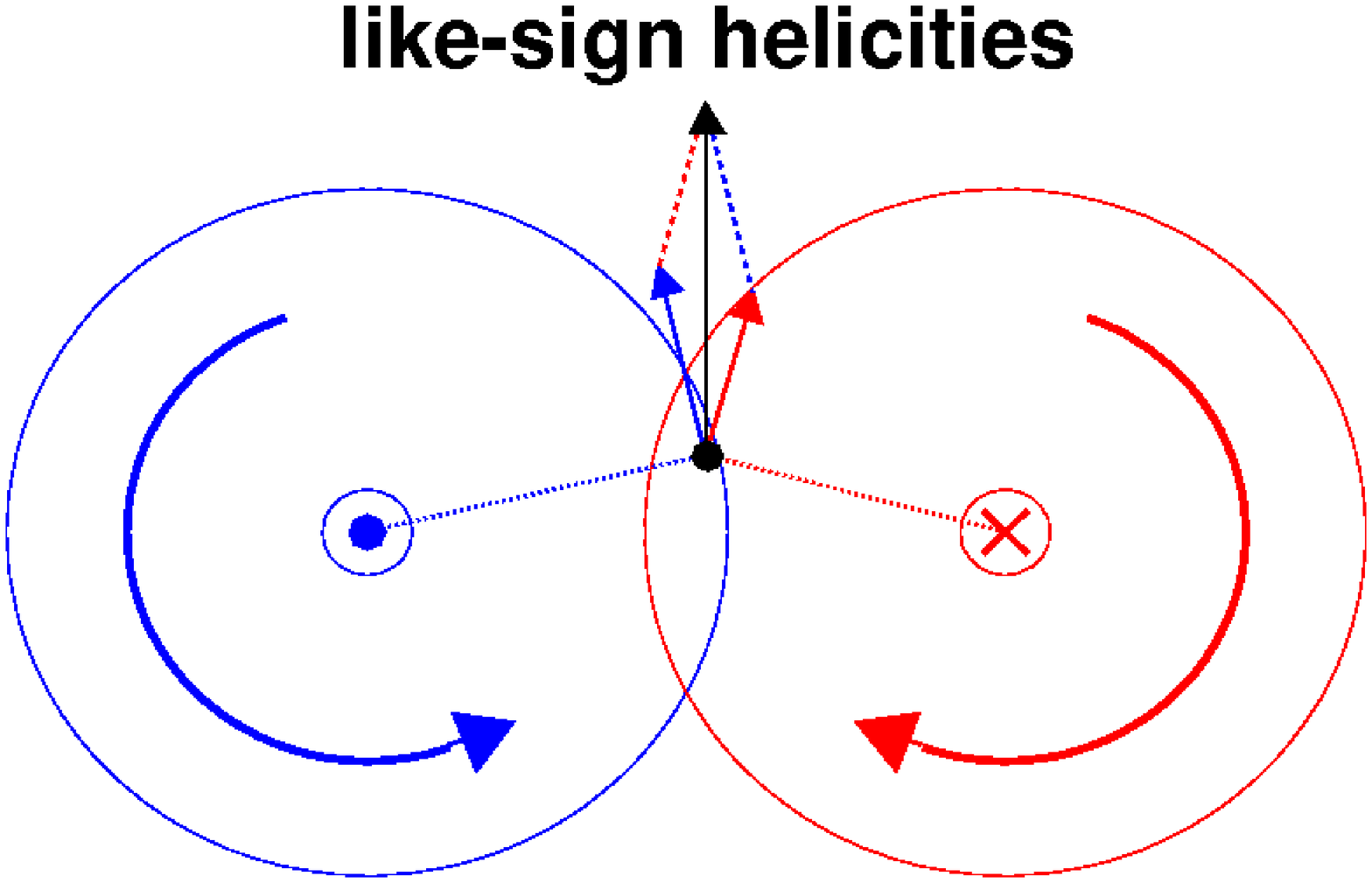}
\includegraphics[width=0.48\linewidth]{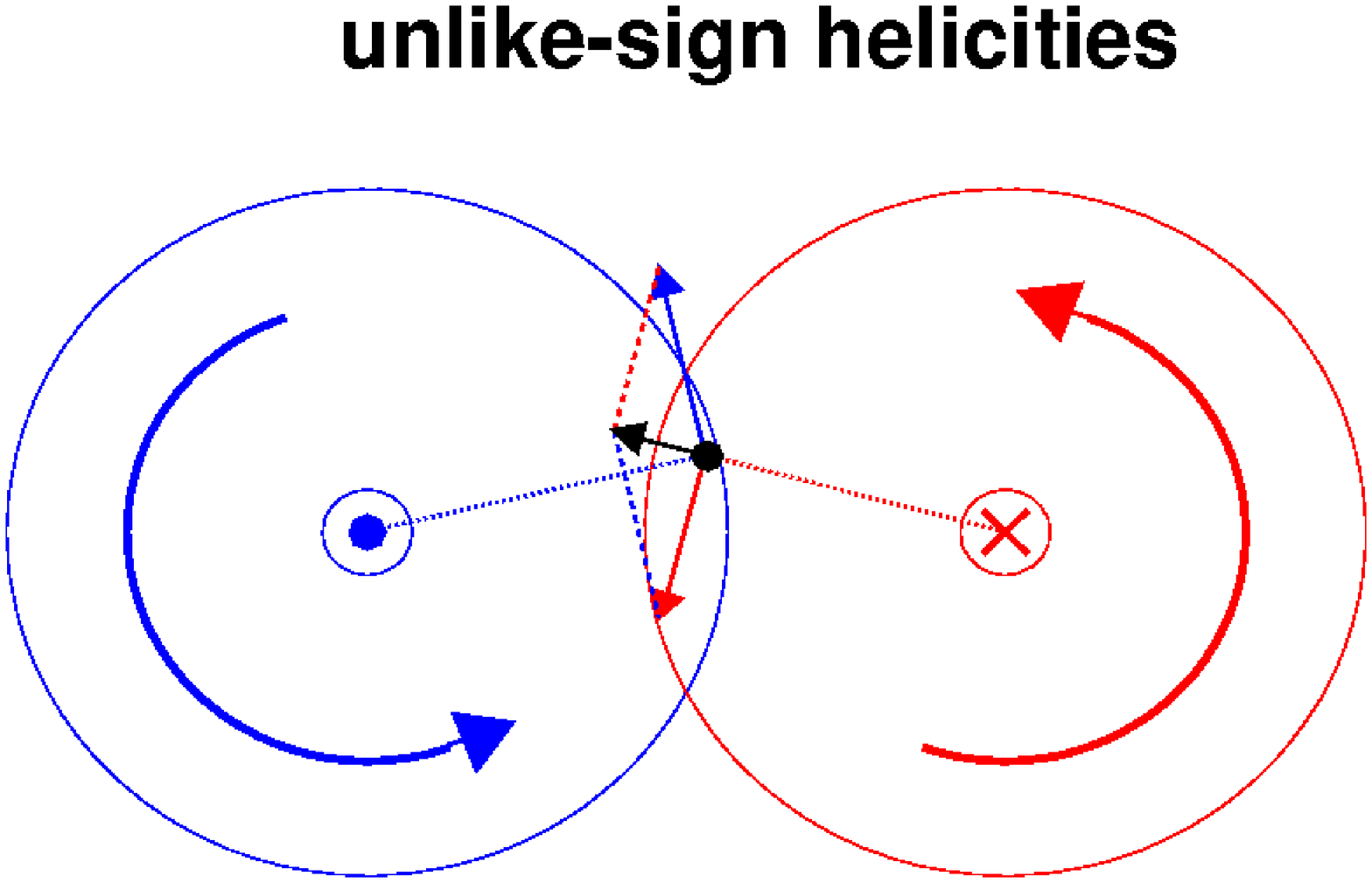}
\includegraphics[width=0.48\linewidth]{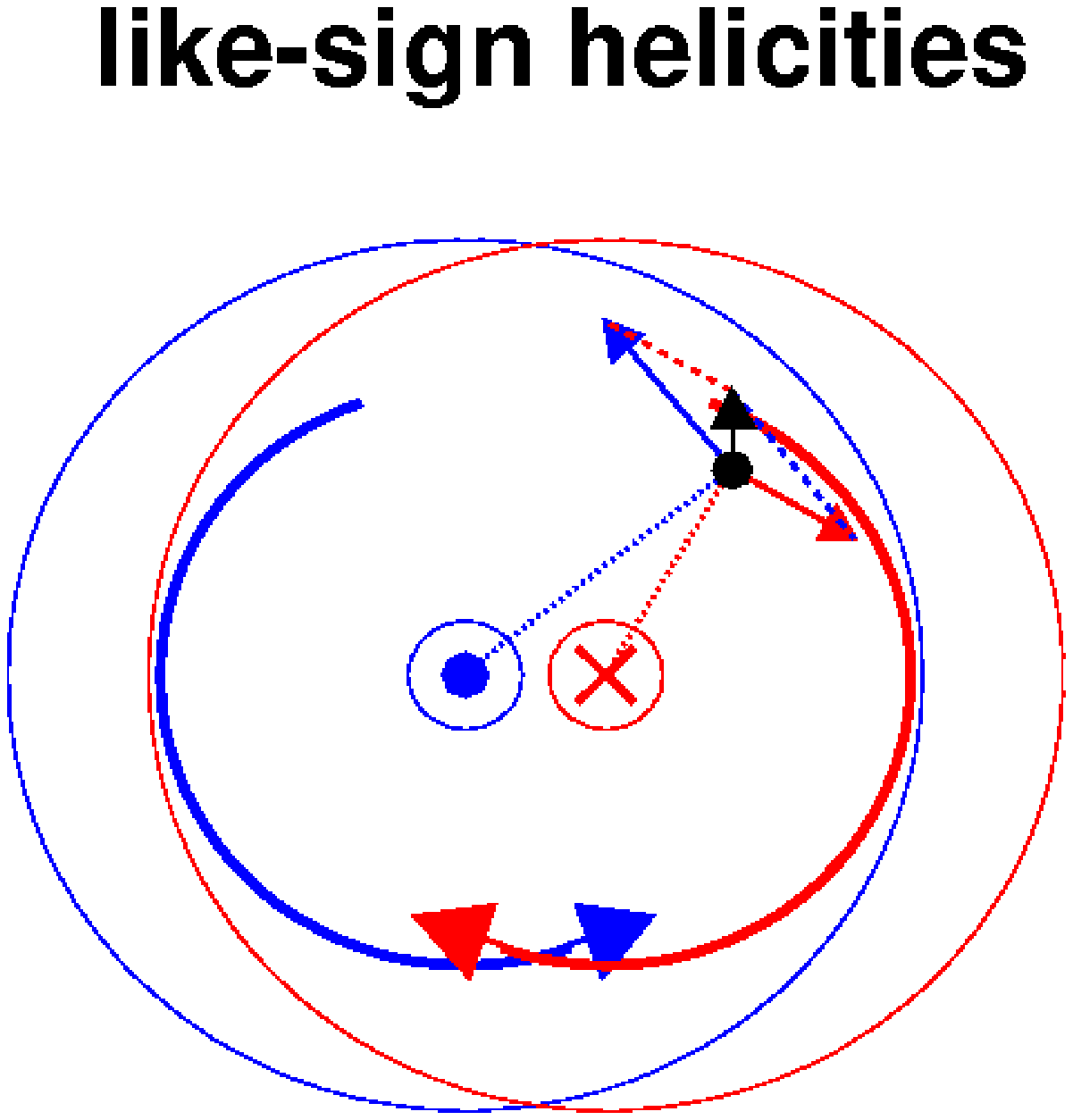}
\includegraphics[width=0.48\linewidth]{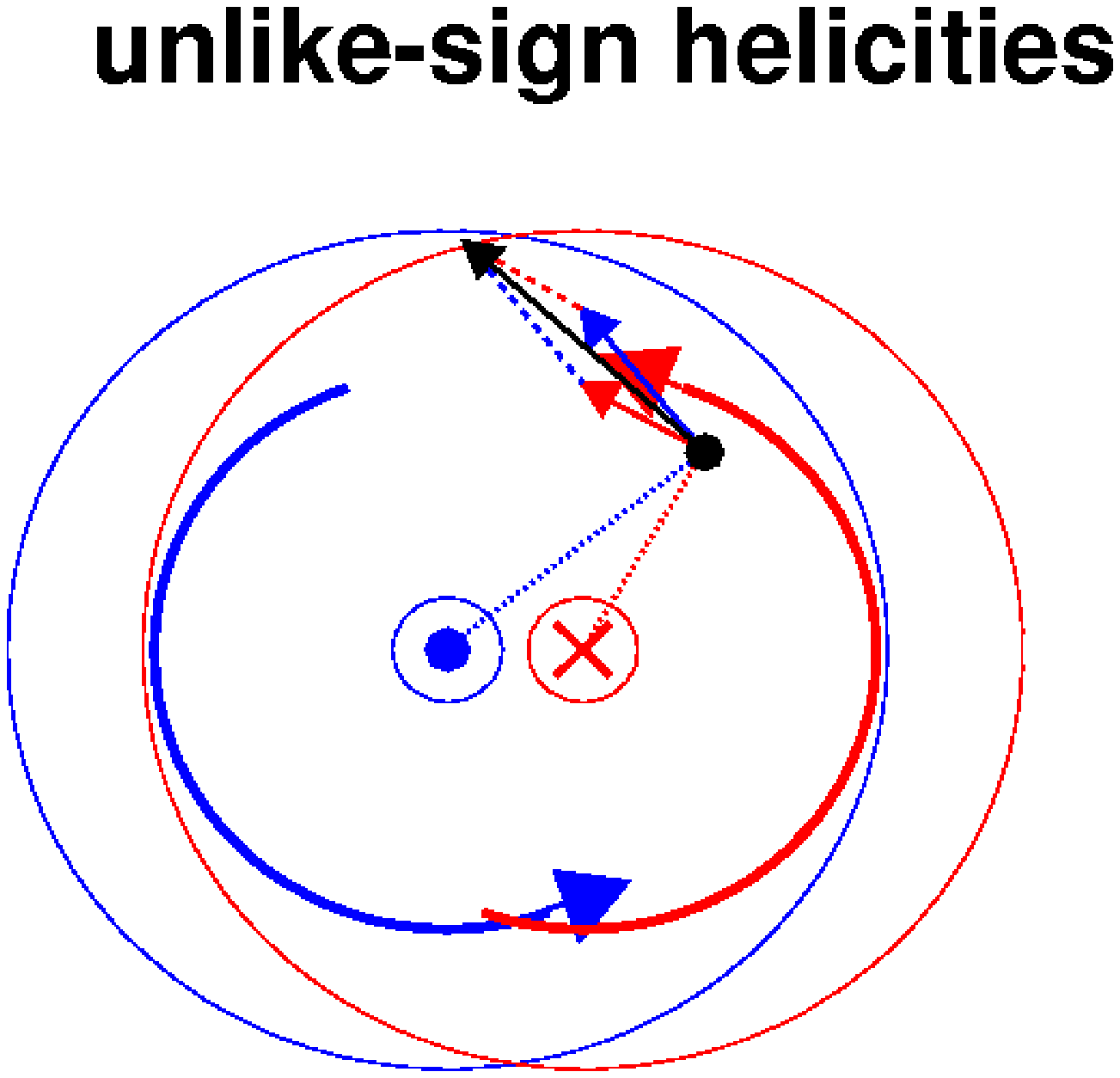}
\caption {\label{fig:mengeffect} (Color online) 
Colliding protons are represented by overlapping circles, with 
proton momentum designated by the central symbol, and spin 
direction designated by the clockwise or counter-clockwise 
arrows. A positive correlation between parton transverse 
momentum and proton spin has been assumed. For a like-sign 
helicity combination (positive on positive, left panels), the 
transverse momenta of the rotating partons add for peripheral 
collisions (top left) and result in a net transverse momentum 
of the lepton pair (in the case of Drell-Yan) and mostly 
cancel for small-impact-parameter collisions (bottom left). In 
the other helicity combination (unlike-sign), the opposite 
effect is seen, i.e., peripheral collisions lead to 
cancellations of the transverse momentum (top right), while 
small-impact-parameter collisions give a larger net transverse 
momentum (bottom right).}
\end{figure}

While the majority of work related to investigating OAM of 
the partons within the nucleon has taken place since the 
1990s, an early theoretical discussion of orbital angular 
momentum inside hadrons was published by Chou and Yang in 
1976~\cite{ref:CHOUYANG}, describing the ``hadronic matter 
current" inside a polarized hadron. After the EMC 
result~\cite{ref:EMC}, Meng et al.~\cite{ref:MENG} built upon 
these semi-classical ideas and proposed two experiments to 
access rotating constituents in the nucleon, one in 
semi-inclusive deep inelastic scattering of unpolarized 
leptons on transversely polarized protons, and the second in 
the measurement of the net transverse momentum of Drell-Yan 
pairs in collisions of longitudinally polarized protons. The 
latter lays the theoretical basis for this analysis:  if the 
transverse momentum of the partons in the initial state is 
correlated with the (longitudinal) spin direction, then hard 
collisions involving these circulating partons will lead to 
final states with a net transverse momentum $p_T$ with 
magnitude dependent upon the relative orientation of the spin 
directions and the impact parameter of the collision, as can 
be seen in Fig.~\ref{fig:mengeffect}.

For a particular helicity combination, e.g., positive on 
positive, the transverse momenta of the rotating partons add for 
peripheral collisions and give a net transverse momentum to the 
lepton pair (in the case of Drell-Yan). For 
small-impact-parameter collisions in the like-helicity 
combination, the helicity-correlated transverse momenta of the 
partons mostly cancel. In the other helicity combination 
(unlike-sign), the opposite effect is seen, i.e., peripheral 
collisions give a small net transverse momentum, while 
small-impact-parameter collisions give a larger net transverse 
momentum.

The correlation of the parton transverse momentum with the 
orbital angular momentum is expected to depend on the spatial 
position of the parton in the proton.  However, experimentally 
there is currently no technique for determining the impact 
parameter of an inelastic $p+p$ collision, and more 
specifically the spatial location of the parton-parton hard 
scattering within that geometry.  Despite this limitation, 
in~\cite{ref:MENG}, with a rather simple picture of the 
transverse spatial distribution (homogeneous sphere) and 
momentum distribution (rotational momentum, $k_\phi$, 
independent of position inside the proton), it was found that 
approximately half of the maximum effect ($\langle p_T^2 
\rangle_{\rm max} = 4 k_\phi^2$, when the vector transverse 
momenta are exactly aligned) remains after integrating over 
the impact parameter.  This result is based on a 
semi-classical model, with the assumption that all interacting 
partons have the same rotational momentum.  As in the case of 
TMDs, there is at present no well-defined relationship 
between the partonic OAM to which this method could provide 
sensitivity and either the Jaffe-Manohar or Ji decomposition 
of nucleon angular momentum. However, it is interesting to 
note that unlike effects due to the Sivers 
TMD~\cite{ref:Brodsky}, in the semi-classical model in which 
the current analysis is framed, the effect discussed below 
does not require an initial- or final-state interaction to 
generate a non-zero effect.

\section{Drell-Yan vs. Jet \kt}

Here, we propose to probe the spin-correlated transverse momentum 
of partons within longitudinally polarized protons involved in 
hard collisions leading to jet-like events at the PHENIX 
experiment at RHIC.  However, in PHENIX, due to our limited 
acceptance (in the central region, $ \Delta\phi = \frac{\pi}{2} 
\times 2$ and $\vert\eta\vert < 0.35$~\cite{ref:PHENIXNIM}), 
we 
do not reconstruct the true jet kinematics to access the jet 
transverse momentum.  An alternative method has been developed 
\cite{ref:ppg029} that examines the dihadron azimuthal angle 
correlation to extract the average parton transverse momentum, 
\ktrms, on a statistical basis for two subsets of the data, 
like-helicity collisions and unlike-helicity collisions, which 
can then be compared as a measure of the helicity dependence of 
the net interacting parton transverse motion.

Since, in contrast with the Drell-Yan experiment proposed in 
\cite{ref:MENG}, we deal here with hadronic final states, there 
could in principle be spin-dependent contributions to the 
measured \ktrms~which are not related to the initial partonic 
transverse momentum.  The measured dihadron transverse momentum, 
$\langle p_{\rm out}^2 \rangle$, is a convolution of the measured 
fragmentation transverse momentum $\langle \jt^2 \rangle$ and the 
extracted partons' transverse momenta $\langle k_T^2 \rangle$.

With the factorization ansatz for the mean $p_T^2$ of the 
scattered partonic pair presented in~\cite{ref:ppg029},

\begin{equation}\label{eq:factor}
\frac{\langle p_T^2 \rangle_{\rm pair}}{2} = \langle k_T^2 \rangle = \langle k_T^2 \rangle ^{\it I} \oplus \langle k_T^2 \rangle ^{\it S} \oplus \langle k_T^2 \rangle ^{\it H}, 
\end{equation}

\noindent where the superscripts {\it I}, {\it S} and {\it H} 
denote intrinsic, soft (one or several soft gluons emitted) and 
hard (NLO) contributions respectively, one might attempt to 
understand the helicity dependence of each term.  The conclusion 
of~\cite{ref:MENG} is that the difference in the intrinsic 
contribution to the mean square parton transverse momentum 
between positive- and negative-helicity protons, $\Delta \langle 
k_T^2 \rangle ^{\it I}$, could be non-zero, since, with a net 
orbital angular momentum, there would be a non-zero helicity 
difference in the vector-summed \kt~of the initial partons.  
$\Delta \langle k_T^2 \rangle ^{\it H}$ could also be non-zero, 
e.g., given a helicity dependence of three-jet events. This 
contribution is theoretically calculable in perturbative QCD, and 
experimentally, contributions from a hard component should be 
accessible by measuring and comparing the spin-dependent $k_T$ 
difference for several center-of-mass energies.  As in QED 
\cite{ref:Yennie}, soft radiation in QCD is independent of the 
polarization of the emitting particle, so the $\langle k_T^2 
\rangle ^{\it S}$ term would not contribute to any spin-dependent 
$\langle k_T^2 \rangle$ difference.

Additionally, since $\langle \jt^2 \rangle $ is used to extract 
$\langle k_T^2 \rangle$ from $\langle p_{\rm out}^2 \rangle$, it 
is important to note that any possible spin dependence of 
$\langle \jt^2 \rangle $ can be measured directly in this 
analysis.

The relationship of a measured \ktrms~difference to a partonic 
orbital angular momentum is non-trivial.  One can attempt to 
relate the spin-correlated parton transverse momentum to this 
difference:

\begin{equation}\label{eq:dkt1}
\Delta\langle k_T^2 \rangle ^{I} =  \sum_{i,j} c^{ij}W^{ij} \left\lbrace \langle \vec{k}^i_T \cdot \vec{k}^j_T \rangle ^{++} - \langle \vec{k}^i_T \cdot \vec{k}^j_T \rangle ^{+-} \right\rbrace
\end{equation}

\noindent where the sums are over all partons in the colliding 
protons, $c^{ij}$ is the probability of an interaction of the 
$i^{th}$ and $j^{th}$ partons leading to the final state, 
$W^{ij}$ is the (unknown) impact parameter weighting for the 
interaction, and $\vec{k}^i_T$ and $\vec{k}^j_T$ are the 
two-dimensional partonic transverse momenta.  In the case with no 
spin-dependent transverse momentum (no orbital angular momentum), 
the difference between the $++$ (like-helicity) and the $+-$ 
(unlike-helicity) terms vanishes.  The $c^{ij}$ can be calculated 
from parton distribution functions, whereas the $W^{ij}$ may be 
estimated from simulations, given a model for the 
impact-parameter-dependent parton distributions.

It is evident from Eq.~\ref{eq:dkt1} that the mixture of 
initial-state partons leading to a \pizh final state will have an 
impact on the interpretation of the data.  In the central arms of 
PHENIX, where \pizh~correlations are measured, {\sc pythia} 
\cite{ref:PYTHIA} simulations show that $\sim$ 50\% of the events 
leading to \pizh~events are $g-g$ in the initial state at \piz 
transverse momenta below 4 \gevc~(from 4-7 \gevc~the fraction is 
$\sim$ 40\% ), $g-q$ initial states making the majority of the 
remainder.  Only a small fraction of the events are like-flavored 
$q-q$ in the initial state.

It is instructive to examine what happens if the sign of the 
orbital angular momentum is different for different flavors. When 
two partons with the same sign OAM interact in a peripheral 
\pp\ collision, then the transverse momentum adds constructively 
as in the top left panel of Fig.~\ref{fig:mengeffect}, regardless 
of the sign of the OAM.  However, if the two interacting partons 
have opposite sign OAM, then the result would be as in the right 
side of Fig.~\ref{fig:mengeffect}. Therefore, an equal mixture of 
parton interactions with like-sign OAM with unlike-sign OAM would 
result in a zero proton-helicity difference in the RMS transverse 
momenta.  On the other hand, if partons of a certain flavor carry 
no OAM, then interactions involving that flavor would contribute 
nothing to the effect in either helicity case, and only act as a 
dilution to the overall transverse momentum difference.

Given the dominance of gluon scattering for the kinematics of 
this measurement, then, the results could be qualitatively 
interpreted (within the semi-classical model presented) as due to 
a diluted contribution of the gluon orbital angular momentum to 
the partonic \kt.  A more quantitative interpretation would 
require a model for the OAM dependence on flavor and kinematics, 
together with a process and experimental simulation.

\section{\kt~From Dihadron Azimuthal Correlations}

\begin{figure}[bt]
\includegraphics[width=1.0\linewidth]{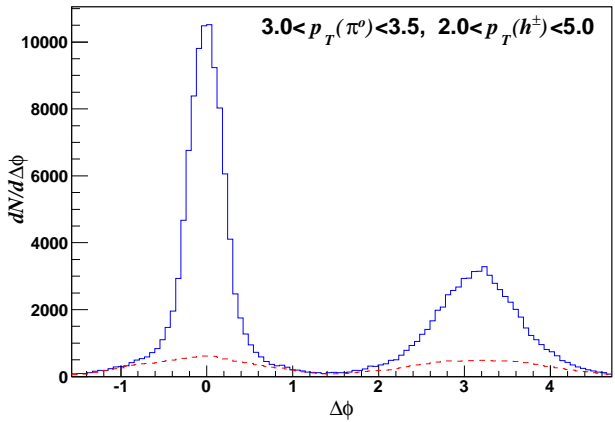}
\caption{\label{fig:dPhiDist} (Color online) 
Azimuthal distributions for real 
(solid curve): $dN_{real}/d\Delta\phi$ and mixed event 
(dashed curve): $dN_{\rm mix}/d\Delta\phi$ pairs.}
\end{figure}

In this analysis we used PHENIX high-\pt~photon triggered data 
from RHIC runs in 2005 and 2006 at \s~= 200 \gev, as has 
previously been published for the PHENIX \piz~cross section 
asymmetry (\all) analysis 
\cite{ref:PHENIX_pi0_all2,ref:PHENIX_pi0_all3} with integrated 
luminosities of 2.5~pb$^{-1}$ and 6.5~pb$^{-1}$ respectively. 
Neutral pions were selected from photon pairs falling in the 
invariant mass region within $M_{\piz} \pm 2.0\sigma$.  The 
signal-to-background ratio for the \piz 's in the range 
$p_{T}^{\piz} \equiv p_{Tt} > 3.0 \gevc$ is above 15.

The azimuthal correlation function is obtained by measuring the 
distribution of the azimuthal (around the beam axis) angle 
difference, $\Delta\phi = \phi_t-\phi_a$, between a \piz 
~(triggered particle) and a charged hadron (associated particle). 
The data is analyzed in eight bins of \piz~transverse momentum 
from $2.0 \gevc < p_{Tt} < 10.0 \gevc$, and the associated 
charged hadron transverse momentum $p_{T}^{h^{\pm}} \equiv 
p_{Ta}$ bin is selected to be within $2.0 \gevc < p_{Ta} < 5.0 
\gevc$ throughout this analysis.  Whenever a \piz~is found in the 
event, the {\it real} ($dN_{real}/d\Delta\phi$) and {\it mixed} 
($dN_{\rm mix}/d\Delta\phi$) distributions are accumulated. The 
mixed event distribution is applied as a correction factor to 
account for the limited PHENIX acceptance.  Mixed events are 
obtained by pairing a \piz~taken from a dihadron event with many 
charged hadrons taken from different events, randomly selected 
from a minimum bias data set (no high-\pt~photon required) 
without regard to helicity. The mixed event distribution is kept 
the same for both helicity combinations. 
Figure~\ref{fig:dPhiDist} shows the real and mixed event 
distributions for $3.5 \gevc < p_{Tt}(\piz) < 4.5 \gevc$.

The fragmentation transverse momentum \jtrms~and the partonic 
transverse momentum \ktrms~are related to the widths of the two 
peaks in the correlation function; around $\Delta\phi=0$ degrees 
to obtain $\sigma_{near}$, and around $\Delta\phi=180$ degrees to 
obtain $\sqrt{\langle p^2_{\rm out}\rangle}$ (the RMS transverse 
momentum of the charged hadrons with respect to the \piz's).  
The raw $dN_{real}/d\Delta\phi$ distribution is fit with the 
following function to obtain $\sigma_{near}$ (based on a 
near-side Gaussian) and $\sqrt{\langle p^2_{\rm out}\rangle}$ 
(based on a more complicated away-side functional form, as 
derived in~\cite{ref:ppg029}):

\begin{widetext}
\begin{equation}
\frac{dN_{real}}{d\Delta\phi} = \frac 1 N \frac{dN_{\rm mix}}{d\Delta\phi} \cdot
\left(C_o + C_1\cdot Gaus(0,\sigma_{near}) + C_2 \cdot \frac{dN_{far}}{d\Delta\phi}
\bigg\arrowvert^{3\pi/2}_{\pi/2}\right)
\end{equation}

\noindent where

\begin{equation}
\label{eq:dn_far}
\frac{dN_{far}}{d\Delta\phi}\bigg\arrowvert^{3\pi/2}_{\pi/2} = 
\frac{-p_{Ta} \cos \Delta \phi}
{\sqrt{2\pi\langle p^2_{\rm out}\rangle}
Erf \left(\sqrt 2 p_{Ta}/\sqrt{\langle p^2_{\rm out}\rangle}\right)}
exp \left( -\frac{p^2_{Ta} \sin^2\Delta\phi}{2\langle p^2_{\rm out}\rangle}\right)
\end{equation}
\end{widetext}


To calculate \jtrms~and \ktrms~from the $\sigma_{near}$ and 
$\sqrt{\langle p^2_{\rm out}\rangle}$ values obtained from the 
fit, the following formulae from~\cite{ref:ppg029} are used:

\begin{eqnarray}
\sqrt{\langle j^2_T \rangle} = \sqrt{2}\frac {p_{Tt} \cdot p_{Ta}} {\sqrt{p^2_{Tt} + p^2_{Ta}}} \sigma_{near}\\
\frac{\langle z_t \rangle \sqrt{\langle k^2_T \rangle}}{\hat x_h} =
\frac 1{x_h} \sqrt{ \langle p^2_{\rm out} \rangle - \langle j^2_{Ty} \rangle \left (1 + x^2_h \right )}
\end{eqnarray}

\noindent where $x_h \equiv p_{Ta}/p_{Tt}$, \xhh~is the analogous 
ratio of the partonic transverse momenta, \mzt~is the ratio of 
hadronic to partonic transverse momentum for the trigger \piz , 
and $\sqrt{\langle j^2_{Ty} \rangle} = \sqrt{\langle j^2_T 
\rangle /2}$.

\begin{figure}[tb]
\includegraphics[width=1.04\linewidth]{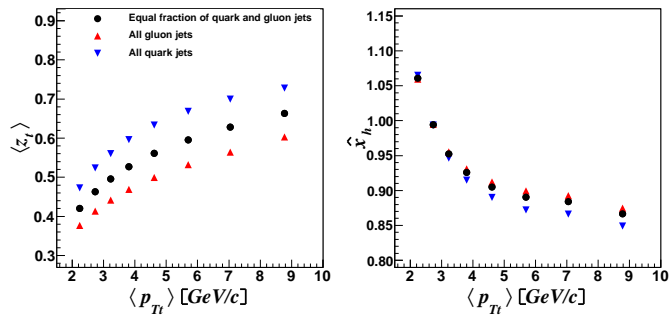}
\caption{\label{fig:xh_zt} (Color online)
Values of derived \mzt~and \xhh~as  explained in the text.}
\end{figure}

Figure~\ref{fig:xh_zt} and Table~\ref{tab:jt_kt} show the 
derived values of \mzt~and \xhh, which were determined 
through an iterative process using a combined analysis of 
the measured \piz~inclusive and associated spectra using jet 
fragmentation functions from LEP 
\ee~measurements~\cite{ref:DELPHI,Alexander:1995bk}, as in 
\cite{ref:ppg029}.  The central values were calculated 
assuming an equal fraction of quark and gluon jets, while 
the systematic uncertainties on \mzt~and \xhh~are estimated 
by taking the RMS spread of the $g-g$, $q-q$ and equal 
fraction initial-state calculations.

\section{Results}

Fits of the $dN_{real}/d\Delta\phi$ distributions were done in 
three ways: 1) all data taken together (summed over spin 
direction), 2) data separated into events from like-helicity 
and unlike-helicity collisions, and 3) the data set randomly 
separated into two sets of approximately equal number of events 
with the like-helicity and unlike-helicity collision type 
assigned randomly.  The first is done as an update to our 
previously published results from 2003 data~\cite{ref:ppg029} 
with higher statistics, but with a slightly different 
associated charged hadron transverse momentum range, and to set 
the baseline for the partonic transverse momentum.  The second 
is the measurement of interest, i.e., the difference in the net 
two-parton transverse momentum in like- versus unlike-helicity 
collisions.  The results of this measurement are to be compared 
to the model result of \cite{ref:MENG}.  The final fitting of 
the randomly assigned helicity combinations is done as a 
measure of the statistical accuracy of the fitting results, as 
explained below.

\subsection{Helicity-Averaged \ktrms~and \jtrms}

The helicity-averaged fit results are enumerated in Table 
\ref{tab:fit_results} for the 2006 data set. The results from the 
2005 data are almost identical, with somewhat larger errors. The 
uncertainties on the fit parameters do not scale with statistics 
across the transverse momentum bins, as the uncertainty on the 
extraction of the width of a Gaussian distribution which is 
superimposed on a constant background does not scale with the 
statistics alone, but also depends upon the width of the 
Gaussian. Since the width of the peaks depends upon the \ptt~bin, 
the uncertainties do not scale with the statistics in each bin. 
Final statistical uncertainties on the fit parameters are 
determined by a statistical technique discussed below.

\begin{figure}[hb]
\includegraphics[width=1.0\linewidth]{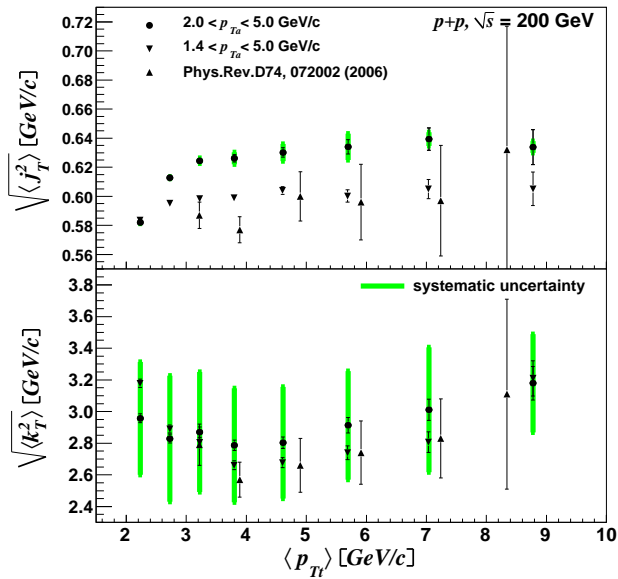}
\caption{\label{fig:JTKT_pT} (Color online)
Helicity-averaged \jtrms~and \ktrms for combined 2005 and 2006 
data.  The systematic uncertainties on the \ktrms~(green bars) 
are due mainly to the systematic uncertainties on the \mzt~and 
\xhh~extractions discussed in the text.}
\end{figure}

The helicity-averaged \jtrms~and \ktrms~results for the 
combined running periods are shown in Table~\ref{tab:syst_err} 
and in Fig.~\ref{fig:JTKT_pT}, where they are also compared to 
the previous results~\cite{ref:ppg029}.  Note in 
Fig.~\ref{fig:JTKT_pT} that the associated charged hadron 
transverse momentum bin is somewhat higher in the current 
analysis, but when checked by lowering the lower limit 
on~\pta, the two results are consistent.

\begingroup \squeezetable
\begin{table}[t]
\caption{\label{tab:jt_kt} 
Calculated values of \xhh and \mzt for the combined 
2006 and 2005 data sets.}
\begin{ruledtabular} \begin{tabular}{ccccc}
&    $p_{Tt}$ & ${\hat x}_h$ & $\langle z_t \rangle$ & \\ 
& $ \gevc $ &              & & \\
\hline
& 2.0-2.5 & $1.061\pm0.003$ & $0.42\pm0.05$ & \\
& 2.5-3.0 & $0.994\pm0.000$ & $0.46\pm0.06$ & \\
& 3.0-3.5 & $0.952\pm0.004$ & $0.50\pm0.06$ & \\
& 3.5-4.2 & $0.926\pm0.008$ & $0.53\pm0.06$ & \\
& 4.2-5.2 & $0.905\pm0.011$ & $0.56\pm0.07$ & \\
& 5.2-6.5 & $0.890\pm0.014$ & $0.60\pm0.07$ & \\
& 6.5-8.0 & $0.884\pm0.013$ & $0.63\pm0.07$ & \\
& 8.0-10.0 & $0.866\pm0.013$ & $0.66\pm0.06$ &
\end{tabular} \end{ruledtabular}
\end{table}
\endgroup

\begingroup \squeezetable
\begin{table}[hbt]
\caption{\label{tab:fit_results} Fit parameters $\sigma_{near}$ 
and $\sqrt{\langle p^2_{\rm out} \rangle}$ extracted from the 
helicity-averaged 2006 data set.  The 2005 results are consistent 
within uncertainties.  The uncertainties on the parameters do not 
scale directly with overall statistics, as discussed in the text.}
\begin{ruledtabular} \begin{tabular}{ccccc}
$p_{Tt} $ & $\langle p_{Tt} \rangle $ & $\langle p_{Ta} \rangle $ 
          & $\sigma_{near}$ & $\sqrt{\langle p^2_{\rm out} \rangle} $\\
$ \gevc $ & $ \gevc $       & $ \gevc $       
          &         & $ \gevc $ \\
\hline
2.0-2.5  & 2.23 & 2.65 & $0.240 \pm 0.001$ & $1.53 \pm 0.02$ \\
2.5-3.0  & 2.73 & 2.67 & $0.226 \pm 0.001$ & $1.42 \pm 0.01$ \\
3.0-3.5  & 3.22 & 2.71 & $0.213 \pm 0.001$ & $1.38 \pm 0.02$ \\
3.5-4.2  & 3.80 & 2.75 & $0.199 \pm 0.001$ & $1.28 \pm 0.02$ \\
4.2-5.2  & 4.61 & 2.80 & $0.187 \pm 0.001$ & $1.18 \pm 0.02$ \\
5.2-6.5  & 5.70 & 2.86 & $0.174 \pm 0.002$ & $1.09 \pm 0.02$ \\
6.5-8.0  & 7.04 & 2.92 & $0.167 \pm 0.003$ & $1.00 \pm 0.02$ \\
8.0-10.0 & 8.78 & 2.94 & $0.158 \pm 0.004$ & $0.96 \pm 0.03$ \\
\end{tabular} \end{ruledtabular}
\end{table}
\endgroup

\subsection{Helicity-Sorted \jtrms and \ktrms}

The process of extracting \jtrms~and \ktrms~was repeated using 
the two subsets of the data corresponding to collisions involving 
like- and unlike-helicity protons at the PHENIX collision area. 
Since any spin-dependent effects should scale with the 
polarization of each beam, all helicity differences are scaled by 
$\frac{1}{P_{B}P_{Y}}$, where $P_{B}$ and $P_{Y}$ are the 
run-averaged beam polarizations for the two colliding beams, 
``blue" and ``yellow" respectively, and are $P_{B}$ = $0.50$ and 
$P_{Y}$ = $0.49$ in 2005, and $P_{B}$ = $0.56$ and $P_{Y}$ = 
$0.57$ in 2006.  Uncertainties on the polarizations were 
propagated together for the two data sets, resulting in a 4.8\% 
scale uncertainty in the spin-dependent differences.

\begin{figure}[tb]
\includegraphics[width=0.95\linewidth]{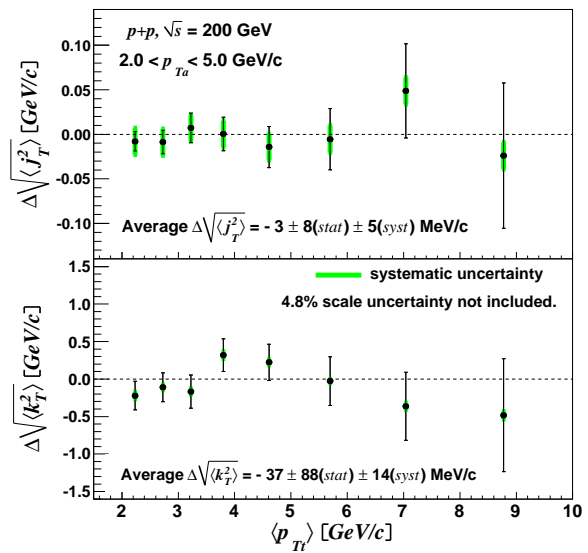}
\caption{\label{fig:d_KT_pT} (Color online)
Difference in \jtrms~(top panel) and \ktrms~(bottom panel) for 
like- minus unlike-helicity combinations.  A systematic 
uncertainty of 4.8\% on the vertical scale due to uncertainty in 
the beam polarizations is not shown. However, this uncertainty 
only affects the relative vertical scale.}
\end{figure}

The helicity-dependent differences for \ktrms~and \jtrms 
~(averaged over the 2005 and 2006 data sets) are shown in 
Fig.~\ref{fig:d_KT_pT}. No \jtrms~difference is observed in any 
\ptt~bin, and if we assume no \ptt~dependence and take the 
average over the \ptt~bins, then the average value of the 
difference in the fragmentation transverse momentum is $\Delta 
\sqrt{\langle j_T^2 \rangle} = -3 \pm 8 ^{\rm stat} \pm 5 ^{\rm 
syst}$ \mevc, consistent with zero.

\begingroup \squeezetable
\begin{table*}[bt]
\caption{\label{tab:syst_err} 
Combined (2005 and 2006) results for \jtrms, \ktrms~and the 
helicity-sorted differences.  First errors are statistical, second 
are systematic.  Statistical and systematic errors are determined 
as described in the text. }
\begin{ruledtabular} \begin{tabular}{cccccc}
$p_{Tt}$ & $\#$ & $\sqrt{\langle j_T^2 \rangle}$ & $\sqrt{\langle k_T^2 \rangle}$ & $\Delta \sqrt{\langle j_T^2 \rangle}$ &  $\Delta \sqrt{\langle k_T^2 \rangle}$ \\ $ \gevc $ & $pairs$ & $ \gevc $ & $ \gevc $ & $ \gevc $ & $ \gevc $ \\
\hline
2.0-2.5 & 792579 & $0.582 \pm 0.001 \pm 0.001$ & $2.96 \pm 0.03 \pm 0.36$ &
$-0.008 \pm 0.011 \pm 0.015$ & $-0.22 \pm 0.19 \pm 0.05$ \\ 
2.5-3.0 & 479497 & $0.613 \pm 0.002 \pm 0.001$ & $2.83 \pm 0.03 \pm 0.40$ &
$-0.009 \pm 0.013 \pm 0.015$ & $-0.11 \pm 0.19 \pm 0.03$ \\
3.0-3.5 & 263174 & $0.624 \pm 0.002 \pm 0.002$ & $2.87 \pm 0.03 \pm 0.38$ &
$0.007 \pm 0.016 \pm 0.015$ & $-0.17 \pm 0.22 \pm 0.04$ \\
3.5-4.2 & 180554 & $0.626 \pm 0.003 \pm 0.004$ & $2.79 \pm 0.03 \pm 0.36$ & 
$0.000 \pm 0.019 \pm 0.015$ & $0.32 \pm 0.22 \pm 0.05$ \\
4.2-5.2 & 101313 & $0.630 \pm 0.003 \pm 0.006$ & $2.80 \pm 0.04 \pm 0.35$ &
$-0.014 \pm 0.023 \pm 0.015$ & $0.22 \pm 0.24 \pm 0.04$ \\
5.2-6.7 & 41827 & $0.634 \pm 0.005 \pm 0.009$ & $2.91 \pm 0.05 \pm 0.34$ &
$-0.005 \pm 0.034 \pm 0.015$ & $-0.03 \pm 0.33 \pm 0.02$ \\ 
6.7-8.0 & 17916 & $0.639 \pm 0.008 \pm 0.005$ & $3.01 \pm 0.07 \pm 0.39$ &
$0.049 \pm 0.053 \pm 0.015$ & $-0.36 \pm 0.45 \pm 0.04$ \\ 
8.0-10.0 & 6775 & $0.634 \pm 0.012 \pm 0.004$ & $3.18 \pm 0.11 \pm 0.31$ &
$-0.024 \pm 0.082 \pm 0.015$ & $-0.48 \pm 0.75 \pm 0.05$ \\ 
\end{tabular} \end{ruledtabular}
\end{table*}
\endgroup

As discussed earlier, there is no quantitative expectation in the 
difference in $\Delta \sqrt{\langle k_T^2 \rangle}$, but any 
non-zero measurement can be attributed to a convolution of 
initial and hard scattering effects.  Since no \ptt dependence is 
expected in the model, and the data are consistent with a flat 
distribution, the difference is averaged over \ptt to get $\Delta 
\sqrt{\langle k_T^2 \rangle} = -37 \pm 88 ^{\rm stat} \pm 14 
^{\rm syst}$ \mevc, consistent with zero.

\subsection{Discussion of Uncertainties}

To check for possible systematic errors due to spin-related 
beam properties or efficiencies, the beam polarization signs 
were randomly chosen for each event with an equal probability, 
the $\Delta\phi$ distributions were obtained for the two false 
helicity combinations, and the fit parameters extracted. This 
process was repeated many times, giving distributions of the 
fit parameters that were well fitted with normal 
distributions.  The widths of the fit-parameter distributions 
for the two false-helicity combinations are then related to 
the statistical fluctuations of the fit parameters.  
Comparison of these widths with the errors returned from the 
fit indicated that the errors on the fit parameters were too 
small by a maximum of $\sim 15\%$, especially for \pout~at the 
larger \ptt~bins.


In order to investigate this non-statistical nature of the fit 
parameter errors, a Monte Carlo simulation was employed. Randomly 
created distributions based on the shapes of the real data 
azimuthal distributions as a function of \ptt~were fitted, 
extracting the fit parameters and errors.  This could then be 
repeated many times, after which the widths of the normal 
distributions of the extracted fit parameters were compared to 
the fit errors.  The exact same trend as a function of \ptt~was 
seen in the Monte Carlo - the fit parameter errors were 
underestimated at the larger values of \ptt~by $\sim 15\%$.  
Since the Monte Carlo is purely statistical, these results 
reflect the true measure of the statistical uncertainty of the 
fit parameters. The statistical uncertainties presented for the 
data are thus those obtained from the spin-randomization 
procedure described above.

The dominant systematic uncertainties were determined from the 
uncertainties in \xhh~and \mzt.  Additionally, the distribution 
of fit parameter values for different mixed-event distributions 
was fitted with a normal distribution, the width of which was 
a measure of the systematic uncertainty of the fit parameters due 
to the detector acceptance correction process.

\section{Discussion}

The smallness of $\Delta \sqrt{\langle j_T^2 \rangle}$ confirms 
the expectation that transverse momentum effects in the 
fragmentation should not be large in processes with a 
longitudinally polarized initial state and thus simplifies the 
interpretation of $\Delta \sqrt{\langle k_T^2 \rangle}$.

Comparing our measured value of $\Delta \sqrt{\langle k_T^2 
\rangle}$ to the calculation of~\cite{ref:MENG},

\begin{equation}\label{eq:MENG_result}
\Delta\langle p_T^2 \rangle \approx 1.9 \langle k_\phi \rangle ^2 
\end{equation}

\noindent yields:

\begin{equation}\label{eq:MENG_relate}
\langle k_\phi \rangle \approx \Delta \sqrt{\langle k_T^2 \rangle} = -37 \pm 88 ^{\rm stat} \pm 14 ^{\rm syst} \mevc,
\end{equation}

\noindent approximately an order of magnitude less than the 
parton intrinsic transverse momentum associated with the 
uncertainty limit. Assuming all contributions to this 
difference come from intrinsic parton motion, and taken 
together with the expected level of contribution from the 
$g-g$ channel and our model assumptions, this could 
qualitatively suggest a small gluon orbital angular momentum 
in a longitudinally polarized proton, integrated over our 
kinematic region.  A more direct connection between this 
measurement and partonic OAM is complicated by the subprocess 
contributions and unknown impact parameter and transverse 
position space weighting of the partons.  In addition, further 
theoretical work is needed to place the model 
of~\cite{ref:MENG} within a rigorous QCD framework. We hope 
that the measurement presented here will serve to encourage 
the theory community to pursue this task.

As discussed in Section \ref{sec:Intro}, since the 1990s there 
has been intense interest in partonic OAM, or more generally, in 
the non-collinear motion of partons within the nucleon, and there 
are several approaches currently being used to attempt to 
increase our understanding of the role that this partonic motion 
plays in nucleon structure, not all of which can be directly 
related to one another.  The measurement presented here, inspired 
by the proposal in~\cite{ref:MENG} for Drell-Yan production 
in 
longitudinally polarized \pp\ collisions, but utilizing a dihadron 
correlation technique, represents a novel experimental approach 
to probing partonic OAM.

A dijet correlation technique in single-{\emph{transversely} 
polarized \pp\ collisions has already been used at RHIC to 
probe the Sivers TMD~\cite{ref:STARSivers}, following a 
proposal in~\cite{ref:AsymJet}.  The current measurement has 
the potential to probe partonic OAM in a longitudinally rather 
than transversely polarized proton.  Dijet and dihadron 
correlation measurements in (polarized) \pp\ collisions provide 
an important tool to investigate the non-collinear motion of 
partons within the (polarized) nucleon, and ideas for expanding 
on the existing techniques would be most welcome.

\section{Summary}

In conclusion, \jtrms~and \ktrms~have been extracted from 
dihadron azimuthal angular correlations in longitudinally 
polarized \pp\ collisions at \s~= 200 \gev.  The helicity 
differences for both quantities are consistent with zero when 
averaged over the \piz~transverse momentum range accessible, 
with a magnitude less than 5\% of the corresponding 
spin-averaged quantities.  Comparison to a similar measurement 
that can be performed on longitudinally polarized \pp\ 
collisions at \s = 500 \gev\ is expected to provide additional 
information regarding hard vs.~intrinsic contributions to the 
measured $\Delta \sqrt{\langle k_T^2 \rangle}$.  The PHENIX 
collaboration collected such a data set in early 2009.  Future 
data at RHIC will increase the statistical significance, and 
upcoming PHENIX upgrades will allow measurements in different 
kinematic regimes, changing the partonic mix probed.  In the 
longer-term future, the accumulation of large luminosities for 
polarized $p+p$~collisions at RHIC should also make possible a 
Drell-Yan measurement, as originally proposed 
in~\cite{ref:MENG}.




\section*{Acknowledgements}   

We thank the staff of the Collider-Accelerator and Physics 
Departments at Brookhaven National Laboratory and the staff of 
the other PHENIX participating institutions for their vital 
contributions.  We also thank D. Sivers, W. Vogelsang, and A. 
Bacchetta for helpful discussions. We acknowledge support from 
the Office of Nuclear Physics in the Office of Science of the 
Department of Energy, the National Science Foundation, a 
sponsored research grant from Renaissance Technologies LLC, 
Abilene Christian University Research Council, Research 
Foundation of SUNY, and Dean of the College of Arts and 
Sciences, Vanderbilt University (U.S.A), Ministry of Education, 
Culture, Sports, Science, and Technology and the Japan Society 
for the Promotion of Science (Japan), Conselho Nacional de 
Desenvolvimento Cient\'{\i}fico e Tecnol{\'o}gico and Funda\c 
c{\~a}o de Amparo {\`a} Pesquisa do Estado de S{\~a}o Paulo 
(Brazil), Natural Science Foundation of China (People's 
Republic of China), Ministry of Education, Youth and Sports 
(Czech Republic), Centre National de la Recherche Scientifique, 
Commissariat {\`a} l'{\'E}nergie Atomique, and Institut 
National de Physique Nucl{\'e}aire et de Physique des 
Particules (France), Ministry of Industry, Science and 
Tekhnologies, Bundesministerium f\"ur Bildung und Forschung, 
Deutscher Akademischer Austausch Dienst, and Alexander von 
Humboldt Stiftung (Germany), Hungarian National Science Fund, 
OTKA (Hungary), Department of Atomic Energy (India), Israel 
Science Foundation (Israel), Korea Research Foundation and 
Korea Science and Engineering Foundation (Korea), Ministry of 
Education and Science, Russia Academy of Sciences, Federal 
Agency of Atomic Energy (Russia), VR and the Wallenberg 
Foundation (Sweden), the U.S. Civilian Research and Development 
Foundation for the Independent States of the Former Soviet 
Union, the US-Hungarian Fulbright Foundation for Educational 
Exchange, and the US-Israel Binational Science Foundation.




\end{document}